%% file: gaugefix.tex
\documentclass[12pt]{JHEP3}
\usepackage{graphicx}
\usepackage{epsfig}
\usepackage{amssymb,amsmath}

\input{newcomm}
\newcommand{\bd}{b^{\dagger}}

\newcommand{\bra}{\langle}
\newcommand{\ket}{\rangle}
\newcommand{\mhinv}{\,m_H^{-1}}
\newcommand{\meff}{m_{\text{eff}}}
\newcommand{\kmax}{k_{\text{max}}}

\title{W and Higgs particle distributions during
electroweak tachyonic preheating}
\author{Jon-Ivar Skullerud, Jan Smit and Anders Tranberg\\
Institute for Theoretical Physics, University of Amsterdam, \\
       Valckenierstraat 65, 1018 XE Amsterdam, The Netherlands.\\
}
\preprint{ITFA-2003-34}

\abstract{We study out-of-equilibrium quasi-particle
distributions of the Higgs and W fields during the
zero-temperature tachyonic electroweak transition that has been
assumed in recent scenarios of baryogenesis. Approximating the
process by a fast quench, we perform classical real-time lattice
simulations in the SU(2)-Higgs model. The emerging quasi-particle
numbers and energies are then used to determine the effective
temperatures, chemical potentials and masses of the particles
shortly after the transition.
}

\begin{document}

\section{Introduction}

An electroweak transition, in which the particles of the Standard Model acquired
their masses,
is assumed to have taken place in the early universe. According to the
standard lore, it was a finite-temperature transition. However, in recent
scenarios of electroweak baryogenesis
\cite{Garcia-Bellido:1999sv,Krauss:1999ng},
this transition is assumed to have taken place at essentially
{\em zero} temperature shortly after low-scale inflation \cite{German:2001tz}
by the effective mass-squared parameter of the Higgs field going negative
(`tachyonic') 
\cite{Copeland:2001qw,Smit:2002yg,Garcia-Bellido:2002aj,Garcia-Bellido:2003wd}.
The resulting spinodal instability has been shown to provide an
effective mechanism of preheating
\cite{Rajantie:2000fd,Rajantie:2000nj,Felder:2000hj,Felder:2000hr,Borsanyi:2002tm}.

These are topics of non-equilibrium field theory.
Quantum fields that are way out of equilibrium need to be treated
non-perturbatively, which is well-known to be a difficult task.
When applicable, classical approximations can be very useful, since they
can be treated by numerical simulation.
The tachyonic electroweak transition is an excellent example, since the
classical approximation is well justified
\cite{Smit:2002yg,Garcia-Bellido:2002aj}.

Whereas results so obtained are within the language of classical
fields, it is desirable to connect with the terminology of kinetic
theory, when appropriate. Particle-distribution functions have
intuitive appeal; they may be describable by time-depend- ent
effective temperatures and chemical potentials. In this paper we
address the problem of extracting particle-distribution functions
from field-correlation functions obtained in classical
approximations.

As is well known, the identification of quasi-particle
distributions in field theory is not unique.
We use a method that was introduced in
\cite{Aarts:1999zn} for fermions and
\cite{Salle:2000hd}
for bosons. This method has also been found
useful in other out-of-equilibrium studies
\cite{Salle:2000jb,Salle:2002fu,Smit:2001qa,Aarts:2001qa,Aarts:2001yn,Cooper:2002qd,Berges:2002wr}, 
and in QCD studies using the
classical approximation applied to the initial stage of heavy-ion collisions
\cite{Krasnitz:2002mn,Krasnitz:2003jw,Lappi:2003bi}.

An important topic in electroweak baryogenesis is the time needed for
the system to reach
approximate thermalization after the transition, and the value of the
corresponding effective temperature.
The thermalization should be fast enough and the temperature low enough
to prevent a possible washout of the generated baryon
asymmetry by sphaleron transtions.
One of the results in this paper is an estimate of this
effective-thermalization time and temperature. However, we shall
also find that a temperature is not sufficient to characterize
the particle distribution and that a substantial chemical
potential is needed as well. A preliminary application to the
present case is in \cite{Skullerud:2002sp}.

In section 2 we introduce the equations of motion and recall the
initial conditions for the tachyonic electroweak quenching transition
\cite{Smit:2002yg}.
Section 3 deals with the definition of distribution functions for the
Higgs and W particles, with more details in the Appendix. In section
4 we present results of numerical simulations: particle numbers and effective
energies and the determination of effective temperatures and chemical
potentials. A discussion of the results is in section 5.

\section{
Tachyonic electroweak transition}

In a tachyonic electroweak transition the effective mass-squared
parameter of the Higgs field in the effective potential is assumed
to change sign from positive to negative. In hybrid inflation
models \cite{Copeland:2001qw,Garcia-Bellido:2003wd}, this is
caused by the coupling of the Higgs field to the inflaton field.
In a first exploration we assume the transition to be dominated by
the dynamics of the SU(2) gauge-Higgs sector of the Standard
Model, and model the transition by a quench. At the electroweak
energy scale the expansion rate of the universe ($O(10^{-5})$ eV)
is negligible
compared to the dynamical time scales of the fields and the
process can be studied in Minkowski space-time.

\subsection{Equations of motion}

The SU(2)-Higgs model is given by the action
\bea
S_{cl}= -\int d^{4}x\left[\frac{1}{2g^{2}}\Tr F_{\mu \nu}F^{\mu \nu}
 +(D_{\mu}\phi)^{\dagger}D^{\mu}\phi
 -\mu^{2}\phi^{\dagger}\phi+\lambda(\phi^{\dagger}\phi)^{2}\right]\, ,
\label{cont_act}
\eea
with
$F_{\mu\nu}=\partial_{\mu}A_{\nu}-\partial_{\nu}A_{\mu}-i[A_{\mu},A_{\nu}]$,
$D_{\mu}\phi=(\partial_{\mu}-iA_{\mu})\phi$. Furthermore,
$A_{\mu}=A_{\mu}^a\tau^a/2$, the
$\tau^a$, $a=1,2,3$ are the Pauli matrices
and $\phi$ is the Higgs doublet
(our metric is $\mbox{diag}(-1,1,1,1)$).
The zero-temperature
Higgs and W masses are given by $m_H^2 = 2\lm v^2 = 2\mu^2$,
$m_W^2 = g^2 v^2/4$, with $v=\mu/\sqrt{\lm}$ the vacuum expectation value
of the Higgs field. The equations of motion are ($m,n=1,2,3$):
\bea
\partial^{2}_{0}\phi &=& D_{n}D_n \phi+\mu^{2}\phi
 -2\lambda(\phi^{\dagger}\phi)\phi,
\label{higgs_eom}\\
\partial_{0}E_{n}^{a}
&=& D_{m}^{ab}F_{mn}^{b}+
\frac{ig^2}{2}[(D_{n}\phi)^{\dagger}\tau^{a}\phi- \phd \ta^a D_n\ph],
\label{gauge_eom}
\eea
with $E^a_n = F^a_{n0}$, $D^{ab}_{n}$ the covariant derivative in
the adjoint representation and we have chosen the temporal gauge,
$A_{0}=0$. The equations of motion for $A^{a}_{0}$
constitute the three Gauss-constraint equations,
which are to be satisfied by the initial conditions. They
are conserved by the equations of motion,
and read 
\bea
D_{n}^{ab}E_{n}^{b}=\frac{ig^2}{2}(\partial_{0}\phi^{\dagger}\tau^{a}\phi
-\phd\ta^a\dnot\ph).
\label{gauss_eom}
\eea

For the numerical simulations the action is discretized on a space-time lattice,
which leads to discretized equations of motion, see e.g.\ \cite{Ambjorn:1990pu}.
More details are given in \cite{Smit:2003??}.

\subsection{Initial conditions}

Before the electroweak transition the system is assumed to be in
the symmetric phase ($\langle \phi \rangle =0$) corresponding to
an effective action in which the term $-\mu^2\phd\ph$ in
(\ref{cont_act}) is replaced by $\mu^2_{\rm eff}\phd\ph$, with
positive $\mu^2_{\rm eff}$. The transition is then caused by
$\mu^2_{\rm eff}$ going through zero and ending up at today's value
$-\mu^2$. We model this process by a quench, in which $\mu^2_{\rm
eff}$ has magnitude $\mu^2$ and flips its sign instantaneously:
$\mu^2_{\rm eff}=\mu^2\to-\mu^2$. This approximation gives
maximal out-of-equilibrium conditions, which we use for testing the
baryogenesis scenario \cite{Smit:2002yg,Smit:2003??}. 
The more gradual transition expected from the coupling to the inflaton
field gives qualitatively similar results
\cite{Garcia-Bellido:2002aj,Garcia-Bellido:2003wd}.

The state just after the quench is unstable (the spinodal
instability), and in the limit
$\lambda\rightarrow 0$, $g^2\rightarrow 0$ it is possible to solve
exactly the time evolution of the momentum modes of the Higgs field
\be
\phi(\veck,t)=\frac{1}{\sqrt{L^{3}}}
\intvecx e^{-i\veck\cdot \vecx}\, \phi(\vecx,t).
\label{def:Fourier}
\ee
Here $L^3$ is the volume of our system with periodic boundary
conditions.

The low momentum modes ($|\veck|<\mu$) initially grow roughly
exponentially until the interaction terms
in the equations of motion become important. Before that happens, the
exponential growth leads to large occupation numbers and to
effectively sharp values of both canonical variables $\ph$ and $\pi=\dnot\ph$,
which
justifies the subsequent use of the classical approximation
\cite{Polarski:1995jg,Smit:2002yg,Garcia-Bellido:2002aj}.
The full nonlinear back-reaction is thus taken into account
without further approximation. This classical approximation should be
reasonable for observables that are dominated by the low momentum modes,
until classical equipartition sets in, which, as we shall see,
does not occur on the time scale of our simulation.

In \cite{Smit:2002yg} it was shown that a consistent way to
initialize the Higgs field (in the initially free-field approximation),
is to generate classical realizations of an ensemble that reproduces the
quantum vacuum correlators in the symmetric phase before the quench, i.e.
\begin{equation}
\langle\phi(\veck)\phi(\veck)^{\dagger}\rangle=
\frac{1}{2\sqrt{\mu^2 + \veck^2}}, \qquad
\langle\pi(\veck)\pi(\veck)^{\dagger}\rangle=
\frac{\sqrt{\mu^2 + \veck^2}}{2}.
\label{eq:higgscorr}
\end{equation}
However, we only initialize the unstable (low momentum
$k<\mu$) modes ($k\equiv|\veck|$). This initial condition scheme is the ``Just
the half'' case of \cite{Smit:2002yg}.

There is an issue concerning how to choose these initial
realizations such that they obey the global Gauss law. This
technical point is treated in Appendix A of our paper
\cite{Smit:2003??} and, for the case of 1+1 dimensions, in
\cite{Smit:2002yg}. Given a realization of the Higgs field with zero total
isospin charges, we can solve the local Gauss-law equations (\ref{gauss_eom})
to find the $E_{i}^{a}$. We set $A_{i}^{a}=0$ initially.
\FIGURE{\epsfig{file=Higgstraj.eps,width=12cm,clip}
\caption{Higgs field expectation values 
$\bra \phi^{\dagger}\phi\ket/v^2$ versus time.
Inset: the long time behavior.}\label{higgstraj}
}

Because only the modes with $k<\mu$ are initialized, we do not
expect problems with lattice artifacts
until the system equilibrates classically. We will see that this
happens on timescales much longer than the ones reached in our
simulations (see also
\cite{Skullerud:2002sp,Moore:2001zf}).

\section{Distribution functions}

Under appropriate circumstances kinetic equations can be derived
in field theory, in which particle numbers constitute a reduced
set of dynamical variables, see, e.g.\
\cite{Blaizot:2001nr} for the case of QCD. Here we are not
concerned with this role of distribution functions, since the
dynamics is treated numerically, but consider them as observables
for studying the preheating process after the quench.

For interacting fields out of equilibrium the definition of local
particle numbers is not unique. At finite (and typically short)
times it is not possible to `go on shell', and there may be
damping and finite-width effects. However, the system may display
effective particle-like behavior in the two-point correlation
functions.

In non-abelian gauge theories there is also the
question of gauge invariance when the chosen two-point functions are not
gauge invariant, as is usually the case. One may render them gauge
invariant by supplying parallel transporters along suitable paths,
as advocated in \cite{Blaizot:2001nr}. This introduces
path-dependence. More generally, there may be field dependence:
different fields with the
same quantum numbers have different correlators and the
resulting particle numbers may or may not depend on the choice
being made.

Part of the ambiguities reside
in the identification of effective-particle energies, the
dispersion relations.
We use a method \cite{Aarts:1999zn,Salle:2000hd}
in which the effective-particle numbers and energies are
determined selfconsistently. Ref.\ \cite{Aarts:1999zn} also
contains a study of the effect of using parallel transporters
and makes a comparison with the Wigner-function approach. See  
also \cite{Berges:2002wr} for further details on the fermionic
case and the relation with conserved charges.

\subsection{Higgs and W particle numbers and effective energies}
\label{sec:eff-n-om}

We are interested in the Higgs- and W-quasi-particle distributions,
which can be obtained from the $\ph$- and $A$-correlation functions in a
suitable gauge. The natural choice of gauge for the Higgs
fields is the unitary gauge,
in which $\ph$ has only one non-zero real component.
For the gauge fields, we will study the particle
distribution in both the unitary gauge and in the Coulomb gauge
$\partial_i A_i=0$.

Writing the Higgs field in the form
\begin{equation}
\phi(\vecx,t) = \frac{1}{\sqrt{2}}\begin{pmatrix}
\vr_2(\vecx,t) + i\vr_1(\vecx,t) \\ \vr_4(\vecx,t)-i\vr_3(\vecx,t) \end{pmatrix}
\label{Hform}
\end{equation}
%
where $\vr_{\al}$, $\al = 1,\cdots,4$ are real,
the unitary gauge is defined by
\be
\vr_a = 0, \;\;\;\; a=1,2,3,
\;\;\;\; h\equiv \vr_4 >0.
\ee
Accordingly, $h=\sqrt{2\phd\ph}$.
The normalization of the fields is chosen such that, in the small amplitude
approximation around a ground state, the fields enter in the kinetic
part of the (in general effective) action with the canonical normalization.
In the present case we have
$S = \intx (1/2) \dnot h \dnot h + \cdots$. After extraction of
the gauge coupling $g$, $A^a_n\to g A^a_n$, the field $A_n^a$ is
also properly normalized. This normalization
criterion can also be applied to composite fields, e.g.\ the rho-meson
field in QCD, provided their effective-action approximation is known.

For simplicity, let us first neglect the coarse graining that
is implicit to the notion of distribution functions, and come
back to this later.
The particle numbers and energies are defined
\cite{Aarts:1999zn,Salle:2000hd}
by analogy with the
free-field expectation values of the equal-time correlators
$\bra\phi\phi\ket$, $\bra\pi\pi\ket$, $\bra AA\ket$, $\bra EE\ket$
(see also the Appendix).
The Higgs particle numbers and effective energies are defined as
\begin{align}
n^H_k(t) &=
 \sqrt{\bra h(\veck,t)h(-\veck,t)\ket_c
\bra\pi_h(\veck,t)\pi_h(-\veck,t)\ket_c} \, ,\\
\om^H_k(t) &=
 \sqrt{\frac{\bra\pi_h(\veck,t)\pi_h(-\veck,t)\ket_c}
      {\bra h(\veck,t)h(-\veck,t)\ket_c}} \, ,
\end{align}
where $\bra\cdots\ket_c$ is the connected two-point function given by
$\bra AB\ket_c \equiv \bra AB\ket - \bra A\ket \bra B\ket$.
We have replaced the `quantum' particle-number combination
$n_k+1/2$ with the `classical' $n_k$, and used spherical
symmetry to write these as a function of $k = |\veck|$.

In a general gauge, we can decompose the gauge field two-point
function in a transverse and a longitudinal part, as
\begin{align}
\bra A^a_i(\veck,t)A^b_j(-\veck,t)\ket &= \dl_{ab}\bigg[
\Big(\dl_{ij}-\frac{k_ik_j}{k^2}\Big)D_T^A(k,t)
+\frac{k_ik_j}{k^2}D_L^A(k,t)\bigg]\,, \\
\bra E^a_i(\veck,t)E^b_j(-\veck,t)\ket &= \dl_{ab}\bigg[
\Big(\dl_{ij}-\frac{k_ik_j}{k^2}\Big)D_T^E(k,t)
+\frac{k_ik_j}{k^2}D_L^E(k,t)\bigg] \,.
\end{align}
The $\dl_{ab}$ reflect the fact that the initial conditions are
isospin symmetric.
In the Coulomb gauge, the gauge potential is purely transverse,
$D_L^A(k,t) = 0$ (but $D_L^E(k,t) \neq 0$),
and the
$n^A_k$ and $\om^A_k$ can be defined analogously to the Higgs case:
\begin{equation}
n^A_k(t) \equiv \sqrt{D^A_T(k,t)D^E_T(k,t)}\,, \qquad
\om^A_k(t) \equiv \sqrt{\frac{D^E_T(k,t)}{D^A_T(k,t)}} \, .
\end{equation}
In the unitary gauge, the free-field correlators are those of a
massive Yang-Mills field
(the derivation is given in the Appendix):
\begin{align}
\bra A_i(\veck,t)A_j(-\veck,t)\ket &=
\Big(\dl_{ij}+\frac{k_ik_j}{m_W^2}\Big)\frac{n_k}{\om_k} \, ,
\label{AA1}\\
\bra E_i(\veck,t)E_j(-\veck,t)\ket &=
\Big(\dl_{ij}-\frac{k_ik_j}{k^2+ m_W^2}\Big)n_k\om_k \,.
\label{EE1}
\end{align}
We decompose this into longitudinal and transverse modes, and arrive
at the expressions we will use to determine longitudinal and
transverse occupation numbers and mode energies in the unitary gauge,
\begin{alignat}{2}
n^T_k &= \sqrt{D^A_T(k)D^E_T(k)}\,, &\qquad
\om^T_k &= \sqrt{\frac{D^E_T(k)}{D^A_T(k)}}\,, \\
n^L_k &= \sqrt{D^A_L(k)D^E_L(k)}\,, &\qquad
\om^L_k &= m_{\rm eff}^{L2}\sqrt{\frac{D^A_L(k)}{D^E_L(k)}} \,.
\label{omL}
\end{alignat}
The transverse case is analogous to the Higgs case.
In the longitudinal case we assumed the form
$\om_k^{L2} = m_{\rm eff}^{L2} + k^2$ and replaced $m_W^2 \to m_{\rm eff}^{L2}$
in (\ref{AA1},\ref{EE1}). Then (\ref{omL}) follows straightforwardly.
Note the inverse dependence on $D^E/D^A$ in $\om^L$.
In practice we analyze the data by first replacing $m_{\rm eff}^L \to m_W$
in (\ref{omL}), and then correct for it.

\subsection{Coarse graining}
\label{sec:volavg}

Consider the problem of defining a position-dependent distribution
function $n(\vecx,\veck,t)$ for a system out of spatial equilibrium.
A natural approach is to consider a region $R(\vecx)$
of size $B$ around the position $\vecx$, e.g.\ a cube of volume
$B^3$, and to focus on this region.
This means restricting the Fourier integrals etc.\ in the formulas in the
previous section to the region $R(\vecx)$.
The size of the region then determines the precision in
momenta and positions of the particles. Similarly, one expects to have to
do some coarse graining in time in order to control the fluctuations in
the energies of the particles. Such time averages were taken in
\cite{Salle:2000hd,Salle:2002fu}, but since we try to follow the
out-of-equilibrium process in time as closely as possible,
we shall not do so here.

For simplicity, consider a scalar field in 1+1 dimensions and let
the localized region $R(x)$ be the interval $(x-B/2,x+B/2)$.
The correlators in momentum space associated with $R(x)$ are then
given by (suppressing the common time label)
\bea
C_{\vr\vr}(x,k) &=& \int_{x-B/2}^{x+B/2} dy\,dz\,
\frac{e^{-iky+ikz}}{B}\, \bra\vr(y)\vr(z)\ket_c,
\nonumber\\
C_{\pi\pi}(x,k) &=& \int_{x-B/2}^{x+B/2} dy\,dz\,
\frac{e^{-iky+ikz}}{B}\, \bra\pi(y)\pi(z)\ket_c,
\eea
from which we obtain the distribution functions in the quantum theory as
\be
C_{\vr\vr}(x,k) = \frac{n(x,k)+1/2}{\om(x,k)},
\;\;\;\;
C_{\pi\pi}(x,k) = (n(x,k)+1/2)\,\om(x,k).
\ee
In the classical approximation the `1/2' is left out.

For a homogeneous system
we can improve statistics by taking a spatial average,
\bea
C_{\vr\vr}(k) &=& \frac{1}{L} \int dx\, C_{\vr\vr}(x,k)
= \sum_p w(p-k)\, \bra\vr(p)\vr(-p)\ket,
\\
w(p-k)&=& \frac{4\sin^2[(p-k)B/2]}{BL(p-k)^2},
\eea
and similar for the $\pi\pi$ correlator ($\vr(p)$ is the
Fourier transform in the total volume,
as in (\ref{def:Fourier})). The weight function $w(p-k)$ is
sharply peaked about $p=k$ and normalized,
\be
\sum_p w(p-k) = 1.
\ee
The particle numbers are obtained {\em after} the spatial averaging,
\be
C_{\vr\vr}(k) = \frac{n_k+1/2}{\om_k},
\;\;\;\;
C_{\pi\pi}(k) =(n_k+1/2)\,\om_k.
\ee
The coarse graining has the effect of smoothing out the correlators in
momentum space. This is a welcome feature, since the initial momentum
modes of the fields are uncorrelated, no matter how large the volume
(their variance is given by (\ref{eq:higgscorr})).

In practise we implement spatial
coarse graining by `binning' our momenta spherically, for example
for the Higgs field:
\begin{equation}
\bra h(\veck,t)h(-\veck,t)\ket \to \frac{1}{N_k}
\sum_{-k-\Delta <|\vecp|<k+\Delta} \bra h(\vecp,t)h(-\vecp,t)\ket,
\end{equation}
where $N_k$ is the number of independent momenta in the momentum bin
labelled by $k$.
In position space this corresponds to spherical shells of thickness
of order $\pi/\Delta$.

\subsection{Early time}
The exponential growth of the particle numbers after the quench was studied in
\cite{Smit:2002yg} in the approximation $\lm = g=0$. For each real
mode $\vr_{\al}$, $\al = 1,\cdots,4$, of the Higgs field, the particle
number in the unstable region ($k < \mu$)
is given by\footnote{Here we correct an error by
a factor 1/2 in eq.\ (3.15) of \protect \cite{Smit:2002yg}.}
\be
n_k^{\al} = \left[\quart + \frac{\mu^4}{4(\mu^4 - k^4)}\,
\sinh^2(2\sqrt{\mu^2 - k^2}\, t)\right]^{1/2}- \half
\approx \quart e^{2\sqrt{\mu^2 - k^2}\, t},
\label{nanalytic}
\ee
where the last form holds for $2\sqrt{\mu^2 - k^2}\, t\gg 1$.
The field correlators are given by
\be
C_k^{\vr_{\al}\vr_{\bt}} = \dl_{\al\bt}\,
\frac{1}{2\sqrt{\mu^2 + k^2}}\left[
1+ \frac{2\mu^2}{\mu^2 - k^2}\, \sinh^2(\sqrt{\mu^2 - k^2}\, t)\right].
\ee
Using the above information
we can estimate the particle number at a time $t_{\rm nl}$
where the neglected nonlinearities may be
expected to stop the exponential growth. We identify $t_{\rm nl}$ with the
time where $\bra\vr_{\al}\vr_{\al}\ket$ (unstable modes only)
reaches the inflexion point $\mu^2/3\lm$ of the Higgs potential, i.e.
\be
\bra\vr_{\al}\vr_{\al}\ket = \sum_{\al}  \int \frac{d^3 k}{(2\pi)^3}\,
C^{\vr_{\al}\vr_{\al}} = \frac{\mu^2}{3\lm}.
\ee
For the parameters ($\lm = 1/9$) used in our simulation this gives
$t_{\rm nl} = 3.32321\, \mu^{-1}$. Then (\ref{nanalytic})
gives for the particle number of each real mode at zero momentum
\be
n_0^{\al}(t_{\rm nl}) = 193.
\label{nHest}
\ee
The distribution at that time is plotted in figure \ref{fig:nom332},
together with
the frequency $\om_k$ \cite{Smit:2002yg}.
\FIGURE{
\includegraphics*[width=7cm]{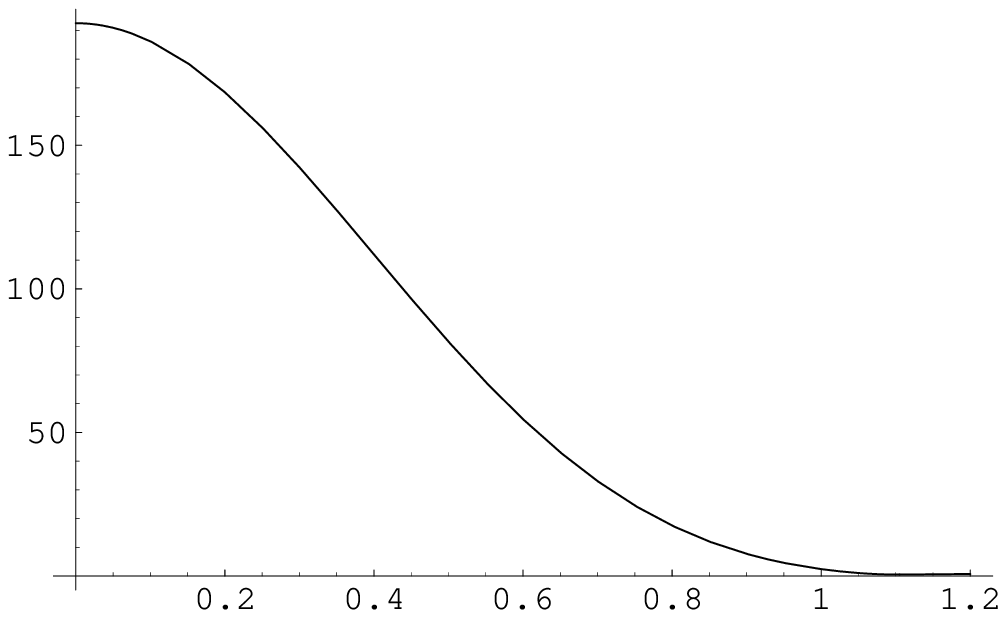}
\includegraphics*[width=7cm]{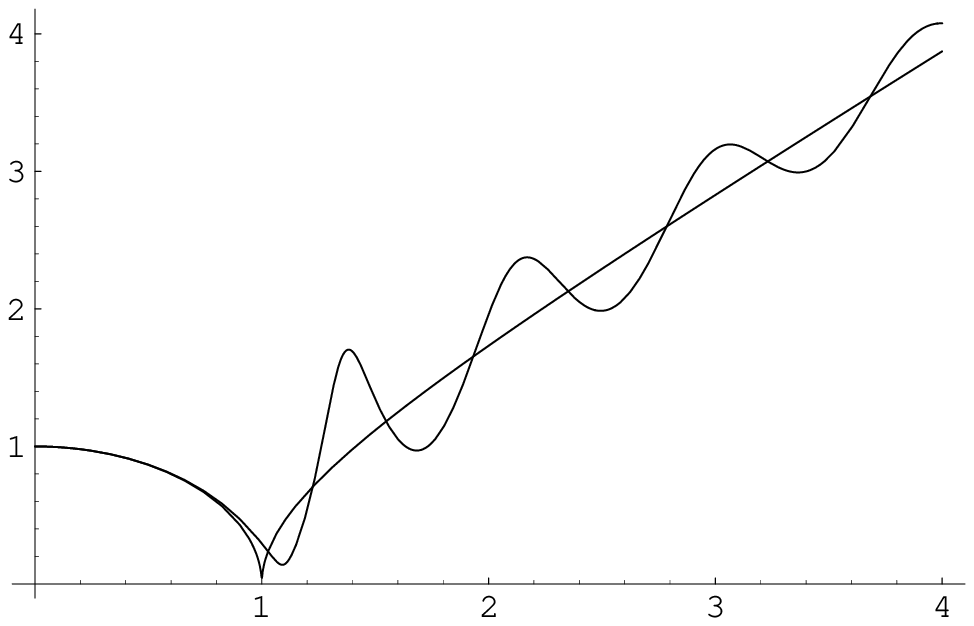}
\caption{Left: $n_k$ for a real Higgs mode
versus $k$.
Right: $\om_k$ and $\sqrt{|\mu^2-k^2|}$ versus $k$.
The units are $\mu = m_H/\sqrt{2}$ and the
time is $t=3.32321\, \mu^{-1} = 4.7\mhinv$. }
\label{fig:nom332}
}
We have not attempted to compute the corresponding quantities for the
radial mode $h = \sqrt{\vr_{\al}\vr_{\al}}$, but
expect them to be similar.

\section{Results}
\label{sec:results}

We have performed simulations with $g=2/3$,
$\lambda=1/9$, giving $m_H^2/m_W^2=2$, and volume $L^3 = 21^3\, m_H^{-3}$.
All results shown will be quoted in units of the zero-temperature
Higgs mass $m_H$.

We used a $60^3$ lattice, with lattice spacing $a=0.35\, m_H^{-1}$.
The maximum momentum in each direction is
$\pi/a = 9.0\, m_H$, but reasonably accurate
continuum behavior
is expected to be limited to the region $|k_i| < 1/a = 2.8\, m_H$.
As is well-known, lattice-artifacts are substantially reduced by
using the corrected momentum $k'_i = (2/a)\sin(a k_i/2)$. In the
following we always use $k'_i$ and drop the prime.

We generated 42 independent
realizations of the initial conditions (\ref{eq:higgscorr}),
and sampled the subsequent time evolution for $tm_H=1$, 2,
\ldots, 12, 20, 30, 40, 50, 100.  The Coulomb-gauge fixing has been
performed using an overrelaxation algorithm 
\cite{Mandula:vs,Cucchieri:1995pn},
stopping when
$L^{-3}\sum_{\vecx,a}|\partial_i A_i^a(\vecx)|^2<10^{-15}$
in lattice units.
This precision has been chosen to ensure that also the low-momentum modes
are tranverse to a high degree of accuracy.

We have averaged nearby momenta as described in
section~\ref{sec:volavg}, within `bins' of size
$2\Delta=0.05\, a^{-1} = 0.0175\, m_H$.
The zero-mode is always in a bin of its own, and will be
treated separately in that it will not be included in any fit
to the data.

\subsection{Particle distributions}
\label{sec:res-n}

\FIGURE{
\includegraphics*[width=12cm]{Phi_n.eps}
\caption{Higgs-particle numbers as a function of $k$.}
\label{fig:nphi}
}
In figure \ref{fig:nphi} we show the particle distribution for the
Higgs fields.  We see that the occupation numbers of the low-momentum
modes increases exponentially up to
$t\approx 6-8\mhinv$, where they start saturating
to an $n_0$ of about 100. 
At this point, the high-momentum modes rapidly
become populated, and after $t\approx 20\mhinv$ the system evolves only
very slowly,
resulting in an approximately exponential distribution.

\FIGURE{
\includegraphics*[width=12cm]{Wtc_n.eps}
\caption{Coulomb-gauge W-particle numbers as a function of
$k$.}
\label{fig:ncoul}
}

\FIGURE{
\includegraphics*[width=12cm]{zeromod.eps}
\caption{Particle numbers of the zero modes as a function of time;
$H$: Higgs; $W_T$: transverse W in unitary gauge; $W_C$: transverse W
in Coulomb gauge; inset: early time.}
\label{fig:zeromod}
}

\FIGURE{
\includegraphics*[width=12cm]{Wtc_n_modes.eps}
\caption{Coulomb-gauge W-particle numbers for the four lowest momentum
  modes, as a function of time.}
\label{fig:n-modes}
}

In figure~\ref{fig:ncoul} we show the Coulomb-gauge W-particle numbers
for all times.
We see the same
qualitative behavior as for the Higgs field, with the occupation
numbers increasing exponentially up to $t\approx6\mhinv$, where they
start saturating to 
$n_0\approx 15$,
followed by a rapid growth in the
high-momentum modes (`sudden up-sweep of the tail')
and
only a slow evolution for $t\gtrsim20\mhinv$.  The
main difference with the Higgs case
is that the initial particle numbers are much lower
--- initially all the energy is in the Higgs fields.  
In figure~\ref{fig:zeromod} we show the time evolution of the particle
numbers $n_0$ of the zero modes.

The exponential growth and subsequent slow evolution can also be seen
by plotting the lowest non-zero momentum modes separately as a
function of time, as in figure~\ref{fig:n-modes}.

\FIGURE{
\includegraphics*[width=7cm]{Wtu_n.eps}
\includegraphics*[width=7cm]{Wlu_n.eps}
\caption{Unitary-gauge transverse (left) and longitudinal (right)
  W-particle numbers as a function of $k$.}
\label{fig:nunit}
}
\FIGURE{
\includegraphics*[width=12cm]{n_WPhi.eps}
\caption{Particle distributions at the latest time, for all four
  particle `species'.}
\label{fig:n-all}
}
Figure~\ref{fig:nunit} shows the W-particle distribution in the
unitary gauge, with the longitudinal and transverse modes plotted
separately.  Note that the longitudinal zero mode is not defined.
The same qualitative behavior as before can be seen.
In this case the initial particle numbers are much larger
and close to those of the Higgs field,
which is natural since the Goldstone modes, which are absorbed into
the unitary-gauge W-fields, are also populated by the initial
conditions (\ref{eq:higgscorr}).

Finally, in figure~\ref{fig:n-all}, we show the particle distribution
for all four `species' (Coulomb-gauge W, transverse and longitudinal
unitary-gauge W, and Higgs) at the latest time, $t=100\mhinv$.  We
see, firstly, that all four have an exponential
fall-off at large momenta.
Secondly, for $k\lesssim 0.7\, m_H$ the occupation number remains
significantly above 1, vindicating our use of the classical
approximation.  Thirdly, the transverse gauge field modes
have almost exactly the same distribution in the unitary gauge as in
the Coulomb gauge.
The distribution of the longitudinal
modes on the other hand deviates
somewhat from that of the transverse ones.

\subsection{Dispersion relation and effective mass }
\label{sec:disp}

\FIGURE{
\includegraphics*[width=12cm]{Wtc_disp.eps}
\caption{Dispersion relation for W-particles in Coulomb gauge.}
\label{fig:disp-coul}
}
In figure~\ref{fig:disp-coul} we show the dispersion relation, $\om^2$
as a function of $k^2$, for Coulomb-gauge W-particles.  For
$t\lesssim20\mhinv$ there is no sensible dispersion relation, as can
be seen from the data for $t=8$; while for $t\gtrsim30\mhinv$ it
approaches the form $\om^2=m_{\text{eff}}^2+ck^2$, with $c\approx1$.
The inset shows the dispersion relation for $t=30,40,50$ and
$100\mhinv$ in a more restricted momentum range.
The data are very well described by a straight line, and for
$t=40, 50$ and $100\mhinv$ they are indistinguishable.  It is striking
that the data turn out to be compatible with
a straight line all the way up to $k^2=40m_H^2$, far into the region
where one would expect lattice artifacts to
dominate and the classical approximation to break down.

\FIGURE{
\includegraphics*[width=7cm]{Wtu_disp.eps}
\includegraphics*[width=7cm]{Wlu_disp.eps}
\caption{W-particle dispersion relation in the unitary gauge, for
  transverse (left) and longitudinal (right) modes.}
\label{fig:disp-unit}
}
In figure~\ref{fig:disp-unit} we show the dispersion relation in the
unitary gauge.  In this case, a particle-like behavior takes
considerably longer to emerge than in the Coulomb gauge: for the
transverse modes the slope is still smaller than 1 (and increasing) at
the latest time, $t=100\mhinv$, while for the longitudinal modes a
curvature remains for $k^2 > m_H^2$.
However,
the intercepts (effective mass-squared) are quite compatible.

\FIGURE{
\includegraphics*[width=12cm]{Phi_disp.eps}
\caption{Dispersion relation for Higgs particles.  Also shown is the
  best fit to a straight line for $t=100\mhinv$.}
\label{fig:disp-phi}
}
In figure~\ref{fig:disp-phi} we show the dispersion relation for the
Higgs particles, for $tm_H=8$, 20, 30, 40, 100.  Here again, we find
that the dispersion relation is stable for $t\gtrsim30\mhinv$.
Evidently,
there is still some remnant of the odd-looking dispersion relation
$\om_k^2 \approx |\mu^2 - k^2|$
during exponential growth at early times, shown in figure \ref{fig:nom332}.
Standard $\om_k^2 = m_{\rm eff}^2 + k^2$-like behavior emerges between
$t m_H = 12$ and 20. Note also that the W-dispersion relation
at $t m_H = 8$ in figures \ref{fig:disp-coul} and \ref{fig:disp-unit} shows
similar behavior: 
the various field modes appear to adjust to each other locally
in momentum space. This can also be seen in the particle numbers.
Similar `local momentum space equilibration' has been
observed and explained in \cite{Salle:2002fu}.

We fit the effective energies to the form $\om_k^2=ck^2+\meff^2$.
The uncertainty in the dispersion relation is estimated by varying the
end-point $\kmax$ of the fitting range and determining the statistical
errors by the bootstrap method.  In figure~\ref{fig:stab-slope} we show
the result of this procedure for the slope of the Higgs dispersion
relation.
\FIGURE{
\includegraphics*[width=12cm]{csmr.eps}
\caption{Slope of the dispersion relation for the Higgs field
 as a function of the maximum momentum in the fit.}
\label{fig:stab-slope}
}
We find that in order to obtain stable values for $c$ and $\meff$, we
need to include points up to $k\approx 2\, m_H$ in our fits, while beyond
this point the fit values do not change significantly.  This was the
case for all the fits we performed, and we have thus chosen
$\kmax=2\,m_H$ when quoting the fit parameters and statistical errors in
table~\ref{tab:dispfits}.
\TABLE{
\begin{tabular}{r|ll|ll|ll}
$t$ & \multicolumn{2}{c}{$W_C$} &\multicolumn{2}{c}{$W_T$}
  & \multicolumn{2}{c}{H} \\ \hline
 & $\meff$ & $c$ & $\meff$ & $c$
 & $\meff$ & $c$ \\ \hline
 30 & 0.56(2) & 0.94(2)  &  0.59(3) & 0.70(5)
  & 0.93(2) & 0.91(3) \\
 40 & 0.64(1) & 0.974(6) &  0.67(1) & 0.69(3)
  & 0.79(3) & 1.02(2) \\
 50 & 0.66(1) & 0.970(7) &  0.67(1) & 0.79(1)
  & 0.84(2) & 1.01(2) \\
100 & 0.68(1) & 0.972(7) &  0.69(1) & 0.828(6)
  & 0.89(2) & 0.98(2)
\end{tabular}
\caption{Effective masses (in $m_H$ units)
and slopes in the dispersion relation.  The
  errors are purely statistical.}
\label{tab:dispfits}
}
For the gauge fields, the intercepts (effective masses) in the two
gauges are compatible, yielding a value
$\meff\approx0.68\, m_H$, close to the zero-temperature value
$0.71\, m_H$.
However, the slope, which in the Coulomb gauge is very close to
1, is considerably lower in the unitary gauge,
although it appears to approach 1 with increasing time.
Quasi-particle behavior appears to take longer to emerge in this gauge.
For the Higgs field, we find an effective mass
$\approx 0.88\, m_H$ at $t= 100 \mhinv$,
which is significantly smaller than the zero-temperature Higgs mass,
although it appears to be increasing with time.
The slope is found to be consistent with 1 from $t=40\mhinv$ onwards.
The effective Higgs mass appears to be increasing with time,
which may be expected from the fact that
$v_{\rm eff}^2 \equiv \bra\phd\ph\ket$ (averaged over an oscillation period)
is still slowly increasing (figure 1).
One would expect a component $v_{\rm eff}/v \approx 0.87$  in
both effective Higgs and W mass ratios, but the effective W mass
somehow seems to have settled from $t = 40 \mhinv$ onwards.

Finally we turn to the longitudinal dispersion
relation $\om_k^L$. In plotting the data in figure
\ref{fig:disp-unit} (right) we used $m_W$ instead of $m_{\rm eff}^{L}$ in
eq.\ (\ref{omL}), and we now consider this point.
Let $\om'_k$ be the frequency defined by (\ref{omL}) with 
$m_{\rm eff}^L \to m_W$.
Then $\om_k^L = (m_{\rm eff}^{L2}/m_W^2) \om_k^{\prime}$ and
$m_{\rm eff}^L = (m_W^2/m')$, where $m' = \om'_0$.
The data for $\om_k^{\prime 2}$ with $k^2/m_H^2 \lesssim 1$ in figure
\ref{fig:disp-unit} (right, $t m_H = 100$) can be fitted well by a
straight line with slope very close to 1, and effective mass
$m^{\prime 2} \approx 0.49\, m_H^2 \approx m_W^2$. It follows that $m_{\rm
eff}^L\approx m_W$ to a good approximation.

The different slopes in the dispersion
relations may be interpretable by simple parameter changes in an
effective quasi-particle lagrangian, but we shall not follow up on this here.


\subsection{Approximate thermalization, temperature and chemical potential}

We will model the particle distribution with a Bose--Einstein (BE)
distribution,
\begin{equation}
n_k = \frac{1}{e^{(\om_k-\mu)/T}-1}.
\label{eq:be}
\end{equation}
It may seem strange to use the BE form instead of the classical
$n_k = T/(\om_k-\mu)$, but this form would not be able to describe
the roughly exponential tail of the $n_k$ data.
We use the BE form simply as a distribution to compare
the data with, in order to extract effective temperatures and chemical
potentials. The BE form may be re-expressed as
\begin{equation}
B_k \equiv \ln\Big(1+\frac{1}{n_k}\Big) = \frac{\om_k-\mu}{T}
\label{eq:invtemp}
\end{equation}
If we have a Bose--Einstein distribution, $B_k$ is
a linear function of $\om_k$, and in a plot of $B$ vs $\om$,
the inverse temperature
can be read off as the slope.
We will therefore refer to such plots as `inverse-temperature plots'.

We will use two methods to determine effective temperatures and
chemical potentials.  The first (method 1) is to perform a straight-line fit to
(\ref{eq:invtemp}), using the data for $n_k$ and $\om_k$.
The second (method 2) is
to take $\om_k$ from the dispersion relation, using the fitted values
for $c$ and $\meff$ determined in the previous section.  We then fit
$n_k$ directly as a function of $\om_k$ to the Bose--Einstein
distribution (\ref{eq:be}).

\FIGURE{
\includegraphics*[width=12cm]{Wtc_nbe.eps}
\caption{Inverse-temperature plot for W particles in Coulomb gauge.}
\label{fig:nbe-coul}
}
In figure~\ref{fig:nbe-coul} we show $B$ as a function of $\om$ for
Coulomb-gauge W particles, for $t=20,30,40,50,100\mhinv$.  For
$\om_k<2.5-3m_H$ the data are compatible with a Bose--Einstein
distribution, with a rather large chemical potential.  At higher
energies, as shown by the inset, there is
some curvature around $\om_k = 4$.
In this region the particle numbers are very small and it is clearly
way beyond the range of validity for the classical approximation.
It is nevertheless
interesting to see that the qualitative behavior of the distribution
is unchanged as we go from the infrared to the ultraviolet.  We also
see that the distribution changes only very slightly with time, with
the effective temperature slowly increasing.  The same qualitative
picture is found also in the unitary gauge.

\FIGURE{
\includegraphics*[width=12cm]{Phi_nbe.eps}
\caption{Inverse-temperature plot for Higgs particles for
  $t=100\mhinv$.  The line is the best fit to a straight line for
  $\om_k<1.6\, m_H$}
\label{fig:nbe-phi}
}
Figure~\ref{fig:nbe-phi} shows $B$ as a function of $\om$ for the
Higgs fields at the latest time.  The statistical errors in $\om$ are 
larger here than for the W fields.
A fit to a straight line through the lowest modes ---
those where $n_k\gtrsim1$ --- deviates from the data at higher $\om$,
and, as seen in the inset, there is a
curvature in the region $\om_k = 2.5 -4$.

\FIGURE{
\includegraphics*[width=12cm]{temp.eps}
\caption{Effective temperature, from inverse-temperature fits, as a
  function of fitting range.}
\label{fig:temp}
}
Figure~\ref{fig:temp} shows the fitted temperature $T$
from the straight-line fits, as a function of the end-point
$\om_{\text{max}}$ of the fitting range.
We have chosen $\om_{\rm max} = 2.1$, which is below the region
where $B_k$ shows curvature in the inserts in figures \ref{fig:nbe-coul}
and \ref{fig:nbe-phi}.

\TABLE{
\begin{tabular}{r|ll|ll|ll}
$t$ & \multicolumn{2}{c}{$W_C$} &\multicolumn{2}{c}{$W_T$}
  & \multicolumn{2}{c}{H} \\ \hline
    & $T_1$ & $\mu_1$     & $T_1$ & $\mu_1$   & $T_1$ & $\mu_1$ \\ \hline
 30 & 0.320(8) & 0.717(12) & 0.24(4)  & 0.73(4)  & 0.23(2) & 1.07(3) \\
 40 & 0.319(3) & 0.772(8)  & 0.268(8) & 0.742(9) & 0.28(3) & 0.94(3) \\
 50 & 0.323(4) & 0.786(9)  & 0.298(3) & 0.758(8) & 0.27(2) & 0.97(3) \\
100 & 0.345(6) & 0.786(9)  & 0.309(4) & 0.781(7) & 0.28(2) & 0.95(3)
\end{tabular}
\caption{Effective temperatures and chemical potentials, from fits to
  $0<\om<2.1m_H$.  The quoted errors are a combination of the
  statistical errors and systematic errors obtained by varying the fit
  range with $\om_{\text{max}}$ between 1.6 and $2.6m_H$.}
\label{tab:nbe-params}
}
The fit values are given in table~\ref{tab:nbe-params}. Since they were
obtained with method 1, we have added the subscript 1 to $T_1$ and $\mu_1$.
The effective
temperature $T_1$ in the unitary gauge turns out to be somewhat
lower than in the Coulomb gauge:
0.31 compared to 0.35\, $m_H$ at the
latest time.  This can be put down to the slope of the dispersion
relation being lower, resulting in a smaller effective energy for the
same mode.  The chemical potentials are consistent between Coulomb and
unitary gauge, giving a value
$\mu\approx0.78\,m_H$. This is higher than the effective mass
$\approx 0.68\, m_H$ in table \ref{tab:dispfits},
which is of course nonsensical, since it would lead to a pole
in $n_k$ for very small $k$.
The effective temperature for the
Higgs is similar to that of the W fields, although slightly lower
$\approx 0.28\, m_H$.
The chemical potential for the Higgs comes out to be $\approx0.95\, m_H$,
again higher than the effective mass $\approx 0.88\, m_H$ in
table \ref{tab:dispfits}.

\FIGURE{
\includegraphics*[width=12cm]{Wtc_nw.eps}
\caption{W-particle number at $t=30,40,50,100\mhinv$ as a function of the
  effective energy $\om$, together with the Bose--Einstein
  distribution (\protect\ref{eq:be}) with parameters from
  table~\protect\ref{tab:nbe-params}. The dotted lines indicate the
variation allowed by the errors in the fitted $T$ and $\mu$. }
\label{fig:nom}
}
Figure~\ref{fig:nom} shows the Coulomb-gauge W-particle number as a
function of the effective energy $\om$, together with the best
inverse-temperature fit to the five lowest-lying modes.  These data do
not appear to be particularly well described by a Bose--Einstein
distribution, largely due to what appears to be a
somewhat erratic
behavior of the effective energy $\om_k$.  Replacing $\om_k$ with
values taken from the fits in table~\ref{tab:dispfits}, we obtain a
much smoother
data plot, as shown in figure~\ref{fig:nomk}.
\TABLE{
\begin{tabular}{r|ll|ll|ll}
$t$ & \multicolumn{2}{c}{$W_C$} &\multicolumn{2}{c}{$W_T$}
  & \multicolumn{2}{c}{H} \\ \hline
 & $T_2$ & $\mu_2$  & $T_2$ & $\mu_2$  & $T_2$ & $\mu_2$ \\ \hline
 30 & 0.50(1)  & 0.62(1)   & 0.33(3)  & 0.64(2)   & 0.45(2) & 0.97(2) \\
 40 & 0.420(7) & 0.698(11) & 0.31(1)  & 0.713(10) & 0.58(6) & 0.83(3) \\
 50 & 0.427(7) & 0.721(9)  & 0.335(7) & 0.719(10) & 0.46(2) & 0.88(2) \\
100 & 0.424(6) & 0.735(9)  & 0.370(5) & 0.733(9)  & 0.38(1) &
0.90(2)
\end{tabular}
\caption{Effective temperatures and chemical potentials, from fits to
  the lowest 5 non-zero modes (method 2).}
\label{tab:nwk-params}
}
\FIGURE{
\includegraphics*[width=12cm]{n_wk_withu.eps}
\caption{Higgs and W-particle numbers at $t=100\mhinv$ as a function
  of the effective energy $\om$ taken from the fitted dispersion
  relation. Also shown is the Bose--Einstein distribution
  (\protect\ref{eq:be}) with parameters from
  table~\protect\ref{tab:nwk-params}.  The zero-modes are outside the
  boundaries of the plot
  (they are given by $(0.68,45(2))$ for $W_C$, $(0.69,52(2))$ for $W_T$
  and $(0.89,12(2))$ for $H$).}
\label{fig:nomk}
}
We fit the 5 lowest-lying modes (for which $n_k\gtrsim0.5$) of these
data to a Bose--Einstein distribution (method 2) and give the results
in table~\ref{tab:nwk-params}.
The fit describes the displayed
data well in the important region where the classical
approximation is supposed to be valid.

Again we find that the effective
temperature for the W particles in the unitary gauge is lower than in
Coulomb gauge, due to the lower slope of the dispersion relation.
On the other hand, the effective Higgs-temperature is falling from
$t=40\mhinv$ onwards and the unitary-gauge W-temperature is rising.
A conversion of energy in the Higgs
to the W may still be going on at the latest time, which is also
suggested by the behavior of the zero modes in figure \ref{fig:zeromod}.

The chemical potentials both for the Higgs and W particles are lower
than those obtained using method 1, reflecting the fact that modes
with small $\om$ have a larger weight.
For the Higgs, $\mu_2$ is compatible with the effective mass,
but for the W it is still larger.
Note that the zero-modes have not been included
in any of the fits.  It turns out that they have occupation numbers
much lower than would be `predicted' by any Bose--Einstein distribution
that would simultaneously fit the other `classical' modes.

This is another indication (in addition to chemical potentials larger
than effective masses) that the BE fit will overestimate the true
distribution for momenta below our lowest finite-volume
momentum $k_{\rm min} = 2\pi /L = 0.30\, m_H$. The flattening of the
distribution in figure \ref{fig:nom332}
as $k\to 0$ appears to be very resilient.

\section{Conclusions}
We have obtained Higgs- and W-particle distributions and energies after
a quenched electroweak transition, when the system is still out of
equilibrium, for the case $m_H = \sqrt{2}\, m_W$. The particle
distributions could be obtained from early times $t=1/m_H$
onwards. On the other hand, the effective energies (frequencies)
suffered much more from fluctuations and they appeared to get
conventional forms only after $t \gtrsim 20\mhinv$. Much more
statistics would be required to go determine them with reasonable
accuracy at earlier times (where they retain the memory of the
instability). The Coulomb-gauge W distribution appears to be
smoother than the corresponding transverse distribution in the
unitary gauge, but the two approach each other and by the time
$100\mhinv$ they are practically indistinguishable. At this stage
the gauge dependence has practically disappeared; also the
longitudinal W modes have settled to nearly the same
distribution. We have carried out a few simulations up to times
$500 \mhinv$, which showed that the distributions change very
slowly after time $100\mhinv$.

The maximum value of the Higgs-particle number
turned out to be smaller than the analytical estimate in (\ref{nHest}),
by a factor of $2$. This mismatch is evidently due to the
fact that with the ``Just a half'' initial conditions,
the interactions (including the Coulomb interaction generated
by the Gauss constraint) are turned on straight away from time zero onwards,
and that the driving power of the instability is damped by sharing
energy with the gauge field.
At low momenta the particle numbers are still large even at $t=100\mhinv$ and
there is no reason to doubt the classical approximation for these modes.
In contrast, the high-momentum modes are still exponentially suppressed
at this time and there is no sign yet of classical equipartition
($n_k\propto 1/k$ for large $k$) in our
simulation. Hence, lattice artifacts are expected to be
small in the low-momentum part of the distributions.

Although we know of no good reason to expect a Bose--Einstein distribution,
this form fitted the data quite well in the important (low momentum) region
(figure \ref{fig:nomk}). The same is true at intermediate momenta,
where the particle numbers are much smaller than one,
albeit with slightly different temperature and chemical potential.
This is fortunate, since it allows for a familiar interpretation of the
results. The final temperatures of the Higgs and W degrees of freedom
turned out to be reasonably close at time $100\mhinv$,
which indicates that the system is near kinetic equilibrium.
In contrast, the rather large chemical potentials show that it is
still far from chemical equilibrium. For very low momenta (lower than our
finite-volume simulation was able to deal with) the BE fit has to break
down because the chemical potentials are slightly
larger than the effective masses.

The temperature in the gauge fields at time $100\mhinv$ is
relatively low, $0.44\, m_H$, which suggest that sphaleron
transitions are highly suppressed. However, such a conclusion
cannot be drawn from the temperature alone. The large chemical
potentials correspond to the fact that there is still quite some
power in the low momentum modes, and sphaleron transitions
might still be non-negligible over a long time span. However, in
the full Standard Model the Higgs and W particles will have
decayed in a few hundred $\mhinv$ and we do not see a problem with
the baryogenesis scenario.

\acknowledgments

This work was supported in part by FOM/NWO.
AT enjoyed support from the ESF network COSLAB.

\appendix
\section{Free-field correlators in Higgs and Coulomb gauge}
\label{appendix}
We derive here the free-field correlators used for guidance in the
definition of the distribution functions in the unitary gauge and
the Coulomb gauge.

In the unitary gauge, expanding around the ground-state
configuration $h = v$, $A_{\mu}^a = 0$, and keeping only terms up
to second order in the fields, leads to the effective free-field
lagrangian
\bea
L_{\rm free} &=&\intvecx
\left[
\half \dnot h\dnot h - \half \dn h\dn h - \half m_H^2 (h-v)^2
\right.\nonumber\\&&\left.\mbox{}
+ \half F_{0n}^a F_{0n}^a - \quart F_{mn}^a F_{mn}^a - \half m_W^2 A_n^a A_n^a
+ \half m_W^2 A_0^a A_0^a\right],
\eea
where we rescaled $A\to gA$ and the nonlinear terms in $F_{\mu\nu}$ should be
dropped. 
To simplify the notation we suppress the common time label $t$ in the following.

The canonical conjugate of the Higgs field is $\pi_h(\vecx) =
\dl L/\dl \dnot h(\vecx)=\dnot h(\vecx)$.
Let $\om_k$ and $n_k$ be defined in terms of the equal-time
correlators, assuming (spatial) translation and rotation
invariance, by
\bea
\bra h(\veck)h(-\veck)\ket_c &=&(n_k+1/2)/\om_k,
\\
\bra \pi_h(\veck)\pi_h(-\veck)\ket_c &=& (n_k+1/2)\, \om_k.
\eea
Since $\pi_h = \dnot h$, the quantity $\om_k$ has the interpretation
of a local (in time) frequency. The meaning of the quantity $n_k$
is elucidated by introducing the annihilation operators
\be
b(\veck) = \frac{1}{\sqrt{2\om_k}}\left[\om_k (h(\veck) - v\dl_{\veck,0}) +
i\pi_h(\veck) \right],
\ee
which satisfy the usual commutation relations with the creation
operators $\bd(\veck)$.
Then,
using the canonical commutation relations we get
\bea
\bra \bd(\veck)b(\veck)\ket &=& n_k  + \frac{i}{2}\bra
\pi_h(\veck) h(-\veck) - \pi_h(-\veck)h(\veck)\ket_c
\nonumber\\
&=& n_k,
\eea
where the last step follows from rotation invariance. So, $n_k$ is
the expectation value of the number operator $\bd(\veck)b(\veck)$.

For the gauge field the free-field effective action of is just the sum
of three actions that are identical in form: one for each value of
the isospin index $a$. For simplicity we suppress the index $a$
and also the index $W$ on $m_W$.
The canonical momenta of the gauge fields are given by
$\Pi^0 = \dl L/\dl \dnot A_0 = 0$, $\Pi_n= \dl L/\dl \dnot A_n =
F_{0n} = (\dnot A_n - \dn A_0) = - E_n$. The field $A_0$ is not an
independent variable, but it follows from the time component of
the field equation $\dmu F^{\mu\nu} - m^2 A^{\nu} = 0$,
i.e.\ $A_0 = \dn \Pi_n/m^2$.
This can be used to express $\Pi_n$ in terms of $\dnot A_n$,
\be
\left(\dl_{mn} - \frac{\dm\dn}{m^2}\right) \Pi_n = \dnot A_m.
\ee
In Fourier space,
\be
\left(\dl_{mn} + \frac{k_m k_n}{m^2}\right) \Pi_n(\veck) = \dnot
A_m(\veck),
\;\;\;\;
\Pi_m(\veck) = \left(\dl_{mn} - \frac{k_m k_n}{m^2 + k^2}\right)
\dnot A_n(\veck)
\ee
The annihilation operators can be defined by analogy to the scalar
case,
\be
a_n(\veck) = \frac{1}{\sqrt{2\om_k}}\left[\om_k A_n(\veck) +
i\dnot A_n(\veck)\right],
\ee
and in this case the canonical commutation relations imply
\be
[a_m(\veck),\ad_n(\veck')] =
\left(\dl_{mn} + \frac{k_m k_n}{m^2}\right)
\dl_{\veck\veck'},
\ee
In terms of spin-polarization vectors
$e^{\mu}(\veck,\lm)$, $\lm = 1,2,3$, satisfying\footnote{These
are the usual orthonormality and completeness
relations of a massive vector field with mass $m$. A realization is given by
$e_n(\veck,\lm)=$ unit vector perpendicular to $\veck$, $\lm=1,2$,
$e_n(\veck,3) = k_n \om_k/k m$. The vectors can be completed with a time
component $e^0(\veck,\lm) = 0$, $\lm = 1,2$,
$e^0(\veck,3) = k/m$, such that they satisfy the standard
relations $e_{\mu}(\veck,\lm)^* e^{\mu}(\veck,\lm')
= \dl_{\lm\lm'}$,
$\sum_{\lm} e^{\mu}(\veck,\lm) e^{\nu}(\veck,\lm) = \et^{\mu\nu} +
k^{\mu} k^{\nu}/m^2$
and $k_{\mu} e^{\mu}(\veck,\lm) = 0$.}
\bea
e_n^*(\veck,\lm)
\left(\dl_{mn} - \frac{k_m k_n}{k^2 + m^2}\right)
e_n(\veck,\lm') &=& \dl_{\lm\lm'},
\\
\sum_{\lm} e_m(\veck,\lm) e_n^*(\veck,\lm) &=&
\left(\dl_{mn} + \frac{k_m k_n}{m^2}\right),
\eea
the usual
annihilation operators for a specific spin state are given by
\bea
a(\veck,\lm) &=& e_m^*(\veck,\lm)
\left(\dl_{mn} - \frac{k_m k_n}{k^2 + m^2}\right)
a_n(\veck),
\\
a_n(\veck) &=& \sum_{\lm} a(\veck,\lm) e_n(\veck,\lm),
\eea
with standard normalization
\be
 [a(\veck,\lm),\ad(\veck',\lm')] =
\dl_{\veck,\veck'}\,
\dl_{\lm\lm'}.
\ee
In a translation- and rotation-invariant state with
\be
\bra \ad(\veck,\lm) a(\veck,\lm)\ket = n_k,
\;\;\;\;
\mbox{independent of $\lm$},
\label{nW0}
\ee
and $\bra a_m(\veck)a_n(-\veck) + \ad_m(-\veck)\ad_n(\veck)\ket = 0$,
we have
\bea
\bra A_m(\veck) A_n(-\veck)\ket &=& \left(\dl_{mn} + \frac{k_m
k_n}{m^2}\right)(n_k + 1/2) \frac{1}{\om_k},
\label{nomW1}
\\
\bra \Pi_m(\veck) \Pi_n(-\veck)\ket &=&
\left(\dl_{mn} - \frac{k_m k_n}{k^2+ m^2}\right)(n_k + 1/2)\,\om_k,
\label{nomW2}
\eea
with $\om_k = \sqrt{k^2 + m^2}$; recall $m = m_W$ in this derivation. 

Conversely, in a more general setting we can follow the same route as
for the Higgs case, define $n_k$ and $\om_k$ by
(\ref{nomW1},\ref{nomW2}) and then derive (\ref{nW0}), which justifies
the interpretation in terms of particle numbers. Evidently, using
rotation invariance makes it possible to avoid having to make the assumption
that $\bra a_m(\veck)a_n(-\veck) + \ad_m(-\veck)\ad_n(\veck)\ket$ vanishes.

Next we consider the Coulomb gauge $\dn A_n^a = 0$. For this case
the free lagrangian is given by
\bea
L_{\rm free} &=& \intvecx \left[\half\dnot h\dnot h + \half
\dnot\vr_a\dnot\vr_a -\half\dn h\dn h - \half \dn\vr_a\dn\vr_a
- \half m_H^2(h^2-v^2)
\right.\nonumber\\&&\mbox{}
+\half \dn A_0^a\dn A_0^a + \half m_W^2 A_0^a A_0^a -m_W A_0^a
\dnot\vr_a
\nonumber\\&&\left.\mbox{}
 + \half \dnot A_n^a\dnot A_n^a -\half\dm A_n^a\dm A_n^a
- \half m_W^2 A_n^a A_n^a
\right],
\eea
where we used $\dn A_n^a = 0$ and the parametrization
(\ref{Hform}) of the Higgs field, $\vr_4 = h$. The canonical momentum of
$A_n^a$ is
$\Pi_n^a = F_{0n}^a = \dnot A_n^a - \dn A_0^a = -E_n^a$.
Again the $A_0^a$ are not independent degrees of freedom but given
by the constraint equation
\be
\dn \Pi_n^a + m_W^2 A_0^a - m_W \pi_a = 0,
\ee
where $\pi_a = \dnot\vr_a$ and $\dn \Pi_n^a= -\dn\dn A_0^a$. So
the $\pi_a$ mix with the longitudinal components
$\Pi_n^{a L}=-\dn A_0^a$. The $\vr_a$ degrees of freedom play the
role of the longitudinal components of the W field, and they can
be shown to have a mass $m_W$. We shall not give further details
as we study those components numerically only in the unitary
gauge where they are included in the gauge field.

The canonical commutation relations of transverse components of
the gauge field have the usual form familiar from Coulomb-gauge
electrodynamics:
$\Pi_n^{aT} = \dnot A_n^a$,
\be
[A_m^a(\veck),\Pi_n^{b T}(-\veck')] = \dl_{ab}\,
\left(\dl_{mn} - \frac{k_m k_n}{k^2}\right)
\dl_{\veck\veck'},
\ee
and the equal-time commutator of $A_n^a$ with $\Pi_n^{aL}$ equals
zero. In an isospin-symmetric, translation- and rotation-invariant
state we can now define $n_k$ and $\om_k$ by
\bea
\bra A_m^a(\veck)A_n^b(-\veck)\ket &=& \dl_{ab}
\left(\dl_{mn} - \frac{k_m k_n}{k^2}\right)
\frac{n_k+1/2}{\om_k},
\\
\bra\Pi_m^{a T}(\veck) \Pi_n^{b T}(-\veck)\ket &=& \dl_{ab}
\left(\dl_{mn} - \frac{k_m k_n}{k^2}\right)
(n_k+1/2)\,\om_k,
\eea
where the frequencies $\om_k = \sqrt{m_W^2 + k^2}$ in the free
case. With the usual transverse polarization vectors
$e_n(\veck,\lm)$, $\lm=1,2$, and annihilation operators
\be
a(\veck,\lm) = e_n^*(\veck,\lm)\frac{1}{\sqrt{2\om_k}}
\left(\om_k A_n^a + i \dnot A_n^a\right),
\ee
it follows that the $n_k$ have the particle number interpretation
\be
\bra \ad(\veck,\lm) a(\veck,\lm) \ket = n_k.
\ee

\end{document}

%% file: newcomm.tex
\newcommand{\intx}{\int d^4 x\,}

\newcommand{\intvecx}{\int d^3 x\,}

\newcommand{\ad}{a^{\dagger}}

\newcommand{\veck}{{\bf k}}

\newcommand{\vecp}{{\bf p}}

\newcommand{\vecx}{{\bf x}}       

\newcommand{\al}{\alpha}
\newcommand{\bt}{\beta}

\newcommand{\dl}{\delta}

\newcommand{\et}{\eta}

\newcommand{\lm}{\lambda}

\newcommand{\ta}{\tau}

\newcommand{\ph}{\phi}
\newcommand{\vr}{\varphi}

\newcommand{\om}{\omega}

\newcommand{\half}{\frac{1}{2}}
\newcommand{\quart}{\frac{1}{4}}

\newcommand{\Tr}{\mbox{Tr}\,}

\newcommand{\dmu}{\partial_{\mu}}
\newcommand{\dm}{\partial_{m}}
\newcommand{\dn}{\partial_{n}}
\newcommand{\dnot}{\partial_{0}}

\newcommand{\phd}{\ph^{\dagger}}

\newcommand{\eela}[1]{\label{#1}\end{equation}}
\newcommand{\eeala}[1]{\label{#1}\end{eqnarray}}
\newcommand{\be}{\begin{equation}}
\newcommand{\ee}{\end{equation}}
\newcommand{\bea}{\begin{eqnarray}}
\newcommand{\eea}{\end{eqnarray}}